\documentstyle[11pt,epsf.tex]{article}

\newcommand{\BE}{\begin{equation}}
\newcommand{\EE}{\end{equation}}
\newcommand{\BA}{\begin{eqnarray}}
\newcommand{\EA}{\end{eqnarray}}
\newcommand{\R}{R_{e^+e^-}}
\newcommand{\lam}{\tilde{\Lambda}}

\begin{document}
\begin{titlepage}
\begin{flushright}
{\small DOE/ER/05096-50}
\vspace{33mm}
\end{flushright}
\begin{center}
{\Huge\bf QCD Perturbation Theory \\
\vspace*{2mm}
          at Low Energies}
\vspace{29mm}\\
{\Large A. C. Mattingly and P. M. Stevenson}
\vspace{18mm}\\
{\large\it
T.W. Bonner Laboratory, Physics Department,\\
Rice University, Houston, TX 77251, USA}
\vspace{30mm}\\
{\bf Abstract:}
\end{center}

We apply the optimization procedure based on the Principle of Minimal
Sensitivity to the third-order calculation of $\R$.  The effective couplant
remains finite, freezing to a value $\alpha_s/\pi = 0.26$ at low energies.
Using Poggio-Quinn-Weinberg smearing we find good agreement between
theory and experiment right down to zero energy.
\end{titlepage}
\setcounter{page}{1}

  Does QCD perturbation theory have anything at all to say about physics
below 1 GeV?  Since the effective couplant:
\BE
a \equiv \frac{\alpha_s(\mu)}{\pi} \approx \frac{1}{b \ln (\mu/\Lambda)}
\EE
(with $b = (33-2N_f)/6$) becomes infinite when $\mu = \Lambda \approx
\mbox{a few hundred MeV}$, it is generally thought that perturbation theory
must completely break down in this region.  However, Eq. (1) is valid
only to leading order:  If higher orders produce a non-trivial zero of the
$\beta$ function at $a=a^*$, then $a$ would remain finite, ``freezing''
to the fixed-point value $a^*$ as $\mu \rightarrow 0$.

  Is this ``fixed-point scenario'' realized?  Seemingly not, since the
calculated coefficients of the $\beta$ function \cite{c2}:
\BE
\label{beta}
\mu \frac{\partial a}{\partial \mu} \equiv \beta(a) =
-b a^2 (1+ca+c_2 a^2+...),
\EE
\BE
\label{c}
c= \frac{153 - 19 N_f}{2(33-2N_f)},
\EE
\BE
\label{c2}
c_2(\overline{\rm MS}) = \frac{3}{16} \frac{1}{(33-2N_f)} \left[
\frac{2857}{2} - \frac{5033}{18}N_f + \frac{325}{54}N_f^2 \right],
\EE
are both positive for fewer than 6 flavours.  However, while $b$ and $c$ are
invariants, $c_2$ depends upon renormalization scheme (RS).  The $c_2$
quoted above is for the ``modified minimal subtraction'' ($\overline{\rm MS}$)
scheme.  In other RS's, though, $c_2$ may be large and negative --- indeed,
one is free to {\it define} the RS such that this is true --- so it is clear
that the fixed-point issue cannot be decided from $\beta$ alone.

  Instead, one should examine the low-energy behaviour of
some physical quantity, such as the QCD correction to the $e^+e^-$
total cross section at a {\it cm} energy $Q$:
\BE
\label{R}
{\cal{R}}(Q) = a(1+r_1 a + r_2 a^2 + ...).
\EE
The issue must also be addressed in the context of a proper resolution of
the RS-dependence problem --- which, in our view, means within the framework
of ``optimized perturbation theory'' (OPT)\cite{OPT}.  This is founded upon
the observation that, while perturbative {\it approximations} to physical
quantities do depend upon RS, the exact result is RS invariant, so the
approximation is only believable where it is stable to small changes in RS.
This ``Principle of Minimal Sensitivity'' is supported by many examples
\cite{OPT} and it has an excellent track record in QCD
phenomenology \cite{Aurenche}.

  An OPT analysis of the fixed-point issue was given in Ref. \cite{KSS}.
It was found that a complete third-order calculation is a prerequisite
for addressing this question (and at that time no such calculation
existed in the QCD case).  To third-order the fixed point is determined
by the equation:
\BE
\label{fix}
\frac{7}{4} + c a^* + 3 \rho_2 {a^*}^2 = 0,
\EE
where $\rho_2$ is the RS-invariant combination:
\BE
\label{rho2}
\rho_2 \equiv r_2 + c_2 -(r_1+\frac{1}{2}c)^2.
\EE
Thus, for $c > 0$, the existence or non-existence of a fixed point (in
third order) is governed by the sign of $\rho_2$.  If $\rho_2$ is positive
then there is no solution for $a^*$, meaning that third-order perturbation
theory breaks down before one reaches zero energy, just as in lower orders.
However, if $\rho_2 $ is negative then a positive root $a^*$ exists,
meaning that OPT yields a finite result down to zero energy.  If the $a^*$
is sufficiently small, then -- with appropriate caveats about nonperturbative
effects -- one can take the perturbative result seriously.

In 1988 a calculation of the third-order coefficient $r_2$ was
reported \cite{old}.  Later, it was found that this result was in error, and
a corrected result was published in 1991 \cite{new}.  The old result gave a
positive $\rho_2$, indicating no hope for perturbation theory at low
energies.  Much worse, though, the $\rho_2$ was so large (65 for 4 flavours)
that even at relatively high energies the third-order corrections were
disconcertingly large.  Several authors noted this as evidence against the
usefulness of ``optimization''\cite{doubters}.  However, with the new
result for $r_2$, the situation is transformed:  the new $\rho_2$ is
negative, and of moderate size ($-12.7$ for 4 flavours)\cite{fntepi}.
At high energies the the net third-order correction is quite small \cite{fnte},
and one finds fixed-point behaviour at low energies \cite{Chyla}.

  In this Letter we report on the results of our investigation of
third-order OPT applied to $\cal{R}$ in $e^+e^-$ and its comparison, using
the Poggio-Quinn-Weinberg (PQW)\cite{PQW} smearing method, to experimental
data in the region $0<Q<6$ GeV\@.  Fuller details will be given in a separate
paper \cite{us}.  We use standard values for the current quark masses
($m_u = 5.6$ MeV, $m_d = 9.9$ MeV, $m_s = 199$ MeV, $m_c = 1.35$ GeV) and a
$\Lambda$ parameter that corresponds to $\Lambda_{\overline{MS}} = 230$ MeV
for 4 flavours \cite{booklet}.

  The matching of $\Lambda$ across flavour thresholds requires comment.
The main effect comes from requiring consistency in the definition of
$\Lambda$ so that $a$ is continuous at a threshold.  The point is well
explained by Marciano \cite{Marciano}, but unfortunately his analysis
uses a truncated expansion of $a$ in powers of $1/\ln(\mu/\Lambda)$.
Avoiding this unnecessary approximation (which would ruin any attempt to
go to low energies) we proceed as follows:  The integrated $\beta$-function
equation can be written as \cite{OPT}:
\BE
\label{intbeta}
\ln(\mu/\lam) = \int^{a(\mu)}_{0} \frac{{\rm d}x}{\beta(x)}
- \int^{\infty}_{0} \frac{{\rm d}x}{\beta^{(2)}(x)}
\equiv \frac{\hat{K}(a)}{b},
\EE
where $\beta^{(2)}(x)$ is the second-order truncation, $-b x^2(1+c x)$, of the
$\beta$ function.  [Our $\lam$ is related \cite{OPT} to the conventional
$\Lambda$ parameter by $\ln(\Lambda/\lam) = (c/b)\ln(2c/b)$.]
Consider a theory with $n+1$ quark flavours, but where we are below the
threshold of one quark.  We could either use Eq. (\ref{intbeta}) with
$N_f = n$, thus defining a $\lam$ for $n$ flavours, or we could use
(\ref{intbeta}) with $n+1$ flavours, breaking the first integration into two
parts:
\BE
\ln(\mu/\lam_+) = \int^{a_{th}}_0 \frac{{\rm d}x}{\beta_{+}(x)}
+ \int^{a(\mu)}_{a_{th}} \frac{{\rm d}x}{\beta_{-} (x)}
- \int^{\infty}_{0} \frac{{\rm d}x}{\beta^{(2)}_{+}(x)},
\EE
where the $+$ and $-$ subscripts mean ``above'' and ``below'' threshold
(i.e., $N_f = n+1$ and $N_f = n$), respectively.  Requiring this to agree
with Eq. (\ref{intbeta}) with $N_f = n$, we obtain the matching condition:
\BE
\ln(\lam_{+}/\lam_{-}) = \frac{\hat{K}_{-}(a_{th})}{b_{-}} -
                         \frac{\hat{K}_{+}(a_{th})}{b_{+}} .
\EE
The energy at which $N_f$ should be incremented is not unambiguously defined.
We elected to make the changeover at the $q \bar{q}$ threshold, so that
$a_{th}$ is the optimized couplant obtained at $Q=2m_q$ \cite{fntemu}.

  The optimization procedure in third order is described in Ref. \cite{OPT},
and details of its implementation are given in Refs. \cite{Chyla,us}.
Briefly, the third-order approximant ${\cal R}^{(3)}$ (defined by truncating
(\ref{beta}), (\ref{R}) after three terms) depends on RS through two
parameters, $\tau \equiv b \ln(\mu/\lam)$ and $c_2$.  The coefficients
$r_1$ and $r_2$ depend on these RS parameters such that the combinations
$\rho_1 \equiv \tau - r_1$ and $\rho_2$ (Eq. (\ref{rho2})) are invariant.
The values of these invariants can be computed from the results obtained
in the $\overline{\rm MS}(\mu = Q)$ scheme \cite{new}:
\BE
r_1(\overline{\rm MS}; \mu \! = \! Q) = 1.9857 - 0.1153 N_f,
\EE
\BE
r_2(\overline{\rm MS}; \mu \! = \! Q) = -6.6368 - 1.2001 N_f - 0.0052 N_f^2
- 1.2395 \: {( {\textstyle \sum} q_i )}^2 \! /
(3 {\textstyle \sum} q_i^2 ),
\EE
with $\tau(\overline{\rm MS}; \mu\!=\!Q) = b \ln (Q/\lam_{\overline{\rm MS}})$,
and $c_2(\overline{\rm MS})$ as given above.  [Note that, for a fixed $N_f$,
$\rho_2$ is a fixed number, while $\rho_1$ is a function of the {\it cm}
energy $Q$.]  The principle of minimal sensitivity picks out an ``optimum''
scheme in which the RS-stability conditions
$\partial {\cal R}^{(3)}/\partial \tau = 0$ and
$\partial {\cal R}^{(3)}/\partial c_2 = 0$, are satisfied.  These equations
together with the definitions of $\rho_1$ and $\rho_2$ and the integrated
$\beta$-function equation (\ref{intbeta}), allow one to solve for the optimized
couplant $\bar{a}$ and the optimized coefficients $\bar{c_2}$, $\bar{r_1}$,
$\bar{r_2}$.  The procedure requires the numerical solution of two
coupled equations.  A good initial guess at the solution can be obtained
from the approximation \cite{PWMR} $\bar{r_2} = - \frac{1}{3} \bar{c_2}$,
$\bar{r_1} =0$ (or, better, $\bar{r_1} = \frac{1}{3}\bar{c_2} \bar{a}$).
At energies below $\lam$ this approximation becomes poor, and the numerical
solution of the equations becomes quite delicate (see \cite{Chyla,us}).
However, in this region one can be guided by the fixed-point
solution \cite{KSS} $\bar{r_2} = - \frac{2}{3} \bar{c_2}$,
$\bar{r_1} = \frac{1}{2} \bar{c_2} a^*$.

  For the effective couplant and for ${\cal R}^{(3)}$ itself we obtain the
results shown in Fig. 1.  Below 300 MeV the effective $\alpha_s/\pi$ is
essentially constant at the value 0.263, which is the fixed-point value
obtained from (\ref{fix}) with 2 massless flavours.  Note that this number is
independent of the $\Lambda$ value \cite{fntecrit}.

  To obtain $\R$, allowing for quark masses, we use the formula \cite{PQW}:
\BE
\R = 3 \sum_i q_i^2 \; T(v_i)[ 1 + g(v_i) {\cal R}],
\EE
where
\BA
v_i  & = & (1-4 m_i^2/Q^2)^{\frac{1}{2}}, \nonumber \\
T(v) & = & v(3-v^2)/2, \\
g(v) & = & \frac{4 \pi}{3} \left[ \frac{\pi}{2v} -
\frac{(3+v)}{4} \left( \frac{\pi}{2} - \frac{3}{4 \pi} \right) \right].
\nonumber
\EA
Except for the use of an effective $N_f$ in obtaining $\cal R$, we are
ignoring mass dependence in the coefficients $r_1, r_2$ because the
calculations have been done only for massless quarks.

  A direct comparison of the resulting $\R$ with experiment is not possible
because nonperturbative effects drastically change the threshold behaviour.
However, PQW \cite{PQW} argue that one can define a suitably smeared
quantity:
\BE
\bar{R}_{PQW}(Q;\Delta) \equiv \frac{\Delta}{\pi} \int^{\infty}_{0}
{\rm d} s' \frac{\R(\sqrt{s'})}{(s'-Q^2)^2 + \Delta^2},
\EE
which is insensitive to nonperturbative effects provided that $\Delta$ is
large enough to smooth out the threshold resonances.  PQW used
$\Delta = 3 \, \mbox{\rm GeV}^2$.  Applying this smearing both to
the perturbative prediction and to the experimental data \cite{data}, we obtain
Fig. 2.  There is some discrepancy in the charm-threshold region, but
below 2 GeV the theoretical and experimental curves are almost
indistinguishable.  If we reduce the amount of smearing, taking
$\Delta = 1 \mbox{\rm GeV}^2$, we obtain Fig. 3.  Here the smearing is not
enough to smooth out the structure in the charm-threshold region, but
is sufficient to smooth out the $\rho, \omega$ and $\phi$ resonances, and
the agreement below 1 GeV remains very good.

   There are uncertainties in the data and in the theoretical prediction,
of course.  On the experimental side there are 10 -- 20\% normalization
uncertainties in the continuum multi-hadron data and about 15\%
uncertainties in the $J/\psi$ and $\psi'$ resonance parameters \cite{data}.
This (and uncertainty in the $c$ quark mass) can account for much of the
discrepancy in the charm region.
However, $\R$ below 1.2 GeV is dominated by $\rho, \omega$ and
$\phi$ resonances, whose parameters are known to 5\% or better.
\cite{booklet,data}  We estimate that the experimental $\bar{R}_{PQW}$ curve
below 1 GeV is trustworthy to roughly $\pm 0.12$ for
$\Delta = 3 \,\mbox{\rm GeV}^2$, and to about $\pm 0.07$ for
$\Delta = 1 \,\mbox{\rm GeV}^2$.

  On the theory side there is, of course, an uncertainty due to the
truncation of the perturbation series.  Above 1 GeV one can see that this
error is small by comparing second- and third-order optimized results.
Below about 1 GeV it is undeniable that the prediction for the net QCD
correction term ${\cal R}$ becomes rather uncertain.
For instance, at $Q=0$ the optimized coefficients
become $\bar{r}_1 = -2.9$, $\bar{r}_2 = 14.6$, with $\bar{a} = 0.26$, giving
a series ${\cal R}(0) = 0.26(1 - 0.76 + 1.01)$ in which the higher-order
terms are comparable to the leading term.  However, at least the signs
alternate, and the corrections are not huge.  We would say that the resulting
${\cal R}$ prediction may well be off by a factor of 2, but  is unlikely
to be off by an order of magnitude.  Thus, although its precise value is
uncertain, ${\cal R}$ near $Q=0$ is rather small ($0.3 \pm 0.3$, say), so
that the uncertainty in $\R$ itself is modest, and smearing further dilutes
the uncertainty.  There are other uncertainties, particularly from the quark
masses.  Altogether we estimate that the theoretical $\bar{R}_{PQW}$ curve
below 1 GeV is trustworthy to roughly $\pm 0.07$ for either $\Delta$ value.

  How significant is the good agreement between the data and the OPT
prediction?  As we just argued, the smeared result is rather
insensitive to ${\cal R}$ because it is small.  However, the important
point is that OPT {\em predicts} that ${\cal R}$ is small {\it down to}
$Q=0$.  One could well have imagined that the QCD correction term
became large at low energies.  As an illustration, consider a
``straw-man'' theory in which ${\cal R}$ is the same as ours
down to 2 GeV, but then continues to rise as $1/\{(9/2) \ln(Q/\Lambda_0)\}$,
with $\Lambda_0 \approx 0.2 \,\mbox{\rm GeV}$, until it reaches some
value $H$, at which value it remains frozen down to $Q=0$.  Such
a theory gives essentially the same $\bar{R}_{PQW}$ prediction as ours if
$H$ is small ($\approx 0.3$), but for larger $H$ it gives a result that
is too large at low $Q$.  Based on our previous estimate of the experimental
uncertainty in $\bar{R}_{PQW}(\Delta = 1 \,\mbox{\rm GeV}^2)$ we can say
that the data implies a limit $H \leq 2$.  Thus, the data can rule out a
large ${\cal R}$ term.

  One could also ask: How predictive is the theory?  How different would the
low-energy data have to be to give a significant disagreement with the
predicted $\bar{R}_{PQW}$ near $Q=0$?  The data below 1 GeV is dominated by
the $\rho$ peak, which, after smearing with $\Delta = 1 \, \mbox{\rm GeV}^2$,
contributes a roughly constant 0.7 -- 0.8  to $\bar{R}_{PQW}$ below 1 GeV\@.
Thus, a 10\% change in the area under the $\rho$ peak would change
$\bar{R}_{PQW}$
by the $\pm 0.07$ estimated uncertainty in the theoretical prediction.
We conclude that perturbative QCD can tell us, at least crudely, the size
of the $\rho$ resonance!

\vspace*{3mm}
\hspace*{-\parindent}{\bf Acknowledgements}

This work was supported in part by the U.S. Department of Energy under
Contract No. DE-AS05-76ER05096.

%
\newpage

\epsfxsize=6.0in
\setlength{\unitlength}{1in}

\epsfxsize=6.0in
\begin{figure}[ht]
 \vskip 1.0 in
   \epsffile[72 216  576 648]{alpha.ps}
\end{figure}
\mbox{}
\vskip -1in
\begin{picture}(6,.001)(0,1.25)
 \put(3.9,4.0){\LARGE ${\cal R}$}
 \end{picture}

\vspace{1.in}
\noindent{\bf Fig. 1.}
\noindent
  The optimized couplant $\bar{a} \; (\,= \alpha_s/\pi)$ obtained from
third-order optimization of ${\cal R}$, where
$R_{e^+e^-} = 3 \sum q_i^2 \, (1+ {\cal R})$ for massless quarks, and
$R_{e^+e^-}$ is the $e^+e^-$ total hadronic cross section normalized by
the cross section to $\mu^+ \mu^-$.  The vertical lines indicate the
$s$ and $c$ quark thresholds.

\newpage
\begin{figure}[htb]
   \epsffile[72 216  576 648]{rb3.ps}
 \label{rb1}
\end{figure}
\noindent{\bf Fig. 2.}
\noindent
  The PQW-smeared $R$ ratio for $\Delta = 3 \: {\rm GeV}^2$, showing the
prediction of ``optimized'' QCD perturbation theory compared to
experiment.  Also shown is the prediction of the naive parton model
(i.e., ${\cal R} = 0$ in Eq. (13)).

\newpage
\begin{figure}[htb]
   \epsffile[72 216  576 648]{rb1.ps}
 \label{rb3}
\end{figure}

\noindent{\bf Fig. 3.}
\noindent
  The PQW-smeared $R$ ratio for $\Delta = 1 \: {\rm GeV}^2$.

\begin{thebibliography} {99}
\bibitem{fntemu} Note that we do {\em not} make the changeover at a
specified $\mu$ because $\mu$ has no physical significance, and depends
on the renormalization convention used to define $\lam$.  Only the
{\em ratio} $\mu/\lam$ has any significance.
\end{thebibliography}
\end{document}